\documentclass[twoside,amsmath,amssymb,nofootinbib,superscriptaddress,notitlepage]{revtex4-1}
\usepackage{graphicx,subfigure}
\usepackage{dcolumn,marvosym}
\usepackage{float,hyperref}
\usepackage{empheq,cleveref}
\usepackage{verbatim}
\usepackage{color}
\usepackage{enumitem}
\usepackage{amsmath}
\usepackage{amsfonts}
\usepackage{graphicx}
\usepackage{bookmark}
\begin{document}

\title{Radiation Gauge in AdS/QCD: inadmissibility and implications on spectral functions in the deconfined phase}
\author{David Dudal}
\email{david.dudal@kuleuven-kulak.be}
\affiliation{KU Leuven Campus Kortrijk - KULAK, Department of Physics, Etienne Sabbelaan 53, 8500 Kortrijk, Belgium}
\affiliation{Ghent University, Department of Physics and Astronomy, Krijgslaan 281-S9, 9000 Gent, Belgium}
\author{Thomas G.~Mertens}
\email{tmertens@princeton.edu}
\affiliation{Joseph Henry Laboratories, Princeton University, Princeton, NJ 08544, USA}
\affiliation{Ghent University, Department of Physics and Astronomy, Krijgslaan 281-S9, 9000 Gent, Belgium}

\begin{abstract}
We point out a subtlety in choosing the radiation gauge ($A_z=0$ combined with the Lorenz gauge) for gauge fields in AdS/QCD for black hole backgrounds. We then demonstrate the effect of this on the momentum-dependence of the spectral functions of the $J/\psi$ vector meson, showing a spreading with momentum and a breaking of isotropy, in contrast to previous results in the literature. We also discuss the dependence on a background magnetic field, following our earlier proposed model.
\end{abstract}

\maketitle
\bookmarksetup{startatroot}

\section{Introduction}
In this note, we would like to discuss in detail the choice of the radiation gauge as it is frequently applied in holographic approaches to QCD. Our specific interest arose from the soft wall model, but the arguments given in the next section do apply to any gauge field in an AdS or AdS black hole background. \\

In the soft wall model, one studies excitations of a gauge field in AdS with action
\begin{equation}
\label{softwall}
S = - \frac{1}{4g_5^2}\int d^{5}x \sqrt{-g} e^{-\Phi} \text{tr}\left[F^{L,MN}F_{L,MN} + F^{R,MN}F_{R,MN}\right],
\end{equation}
for left and right gauge fields $A_L$ and $A_R$. We will denote 5-dimensional indices as $M,N,O,P$ and 4-dimensional (boundary) indices with $\mu,\nu$. The corresponding equations of motion of either gauge field are given by:
\begin{equation}
\partial_{M}\left(e^{-\Phi}\sqrt{-G}G^{MO}G^{NP}(\partial_O A_{P} - \partial_{P}A_{O})\right)=0,
\end{equation}
where the background geometry is either AdS:
\begin{align}
\label{ads}
ds^2 &= \frac{L^2}{z^2}\left(-dt^2 + d\mathbf{x}^2 + dz^2\right)\,,\qquad e^{-\Phi} = e^{-c z^2},
\end{align}
or the AdS black hole:
\begin{align}
\label{adsbh}
ds^2 &= \frac{L^2}{z^2}\left(-f(z)dt^2 + d\mathbf{x}^2 + \frac{dz^2}{f(z)}\right) \,,\qquad
e^{-\Phi} = e^{-c z^2},
\end{align}
with $f(z) = 1 - z^4/z_h^4$ and $z=z_h$ the horizon location. The background includes a dilaton field $\Phi$ whose backreaction on the geometry is assumed to be minor. This model gained popularity to holographically capture important QCD physics, e.g.~because of its correct scaling behavior of the meson spectrum and the reader is referred elsewhere for more motivation and details of this model \cite{Erlich:2005qh,Karch:2006pv,Karch:2010eg}.

As a particular example where the soft model was used to study strongly coupled QCD physics, let us refer to \cite{Fujita:2009wc,Fujita:2009ca} where heavy quarkonia were studied. The authors of \cite{Fujita:2009wc,Fujita:2009ca} suggested choosing a flavor-dependent soft-wall parameter $c$, where the light quarks ($u$, $d$, $s$) are combined into a $SU(3)_L\times SU(3)_R$ soft wall model and the heavy quark of interest (charm in our case) is treated on its own in a $U(1)_L \times U(1)_R$ Abelian model:
\begin{equation}\label{fuj}
S = - \int d^5x \sqrt{-g}~\text{tr}~e^{-c_{\rho}z^2}\mathcal{L}_{light} + e^{-c_{J/\psi}z^2}\mathcal{L}_{charm}.
\end{equation}

Our goal in this work is to compute the momentum-dependence of the $\bar{c}c$ spectral function in this model and to demonstrate that one of the conclusions of \cite{Fujita:2009wc,Fujita:2009ca}, namely that isotropy (rotational invariance) is present in the spectral function, is actually a consequence of a forbidden choice of gauge.

\section{Survey of the radiation gauge in AdS/CFT}
The field $A^{\mu}$ has of course a large gauge redundancy and we will investigate here whether the radiation gauge can always be imposed to begin with. The authors of \cite{Erlich:2005qh,Karch:2006pv} and \cite{Fujita:2009wc,Fujita:2009ca} impose the gauge choice
\begin{equation}
A_{z}=0 , \quad \partial^{\mu}A_{\mu}=0,
\end{equation}
where $\mu=0,1,2,3$, the transverse indices. We first note that, for a diagonal metric that is only warped in the $z$-direction, one readily finds that
\begin{equation}
\nabla_{\mu}A^{\mu} = \frac{1}{\sqrt{-G}}\partial_{\mu}\left(\sqrt{-G}A^{\mu}\right) = \partial_{\mu}A^{\mu} = \partial^{\mu}A_{\mu},
\end{equation}
and hence there is no difference between covariant derivatives and ordinary partial derivatives when computing the divergence of $A^{\mu}$. \\

The main discussion we want to start here is: can this gauge be chosen in the first place?\\
Without loss of generality, we may consider a two-step process to get there.
\subsection{Step 1}
First let $\tilde{A}_{M} = A_{M} + \partial_{M} \chi_1$. We choose $\chi_1$ such that
\begin{equation}
\chi_1 = - \int dz A_z,
\end{equation}
which eliminates $\tilde{A}_z$. In general this $\chi_1$ depends on all 5 coordinates.
\subsection{Step 2}
Next we perform a second gauge transformation $\tilde{\tilde{A}}_{M} = \tilde{A}_{M} + \partial_{M} \chi_2$. Since we want to maintain $\tilde{\tilde{A}}_{z}=0$, we should have $\partial_{z} \chi_2 =0$. Additionally, we want to impose the Lorenz (Landau) gauge in the transverse (4d) dimensions. This can be done by solving the following wave equation for $\chi_2$:
\begin{equation}
\partial^{\mu}\tilde{A}_{\mu} = -\partial^{\mu}\partial_{\mu} \chi_2.
\end{equation}
In terms of ordinary coordinate partial derivatives we have
\begin{equation}
\label{tosolve}
G^{\mu\nu}\partial_{\nu}\tilde{A}_{\mu} = -G^{\mu\nu}\partial_{\nu}\partial_{\mu} \chi_2.
\end{equation}
The tricky part is that this metric still depends on $z$. But we just established that $\chi_2$ does not! In general this equation is hence unsolvable. \\

\noindent Before delving into a more detailed exposition, let us look at an analogous problem in (3+1)d classical electrodynamics: we are inclined to choose the temporal and Coulomb gauge simultaneously (thereby defining the radiation gauge): $A_0 = 0$ and $\partial_i A_i = 0$. The latter leads to the equation for $\chi_2$:
\begin{equation}
\partial_{i}A_{i} = -\partial_{i}^2 \chi_2,
\end{equation}
for time-independent $\chi_2$. This equation is nevertheless solvable: in spite of the time dependence of $A_i$ itself, it holds that
\begin{equation}
\partial_{t}\partial_{i}A_{i} = \partial_i E_i = 0,
\end{equation}
by virtue of Gauss'~law. The above gauge is hence possible by imposing the (sourceless) Maxwell equations. Though, in the presence of sources, the radiation gauge is inadmissible. \\

\noindent In our case, we are interested in the analogous problem of a sourcefree Maxwell field, but this time in a curved background. \\
Maxwell's equations take the form
\begin{equation}
\partial_{M}\left(e^{-\Phi}\sqrt{-G}G^{MO}G^{NP}(\partial_O A_{P} - \partial_{P}A_{O})\right)=0.
\end{equation}
In the gauge $A_z=0$, we consider the $N =z$ component of this equation. For a diagonal metric whose components are independent of $0,1,2,3$, we obtain
\begin{equation}
\label{constraint}
G^{\mu\nu}\partial_z \partial_{\mu}\tilde{A}_{\nu} = 0.
\end{equation}
Taking then the $z$-derivative of (\ref{tosolve}), we obtain
\begin{equation}
(\partial_zG^{\mu\nu})\partial_{\nu}\tilde{A}_{\mu} = -(\partial_z G^{\mu\nu})\partial_{\nu}\partial_{\mu} \chi_2.
\end{equation}

In the particular case of a metric for which all components have the same $z$-dependence, we can factor out all $z$-dependence, and rewrite the above as
\begin{equation}
\hat{G}^{\mu\nu}\partial_{\nu}\tilde{A}_{\mu} = -\hat{G}^{\mu\nu}\partial_{\nu}\partial_{\mu} \chi_2,
\end{equation}
where $\hat{G}$ is independent of $z$. Even though $\tilde{A}$ depends on $z$, the constraint (\ref{constraint}) implies the left hand side is $z$-independent, just as the right hand side. No problems occur and a solution of the above equation in terms of $\chi_2$ is possible. This is the case for AdS spacetimes (\ref{ads}). \\

For the AdS black hole (\ref{adsbh}), the metric has a different $z$-dependence for the $G_{00}$ component compared to the other components. For such a space, we can write equation (\ref{tosolve}) as
\begin{equation}
G^{00}\partial_{0}\tilde{A}_{0} + G^{ii}\partial_{i}\tilde{A}_i= -G^{00}\partial_{0}\partial_{0} \chi_2 -G^{ii}\partial_{i}\partial_{i} \chi_2.
\end{equation}
Dividing by $G^{00}$ and differentiating w.r.t.~$z$ we obtain
\begin{equation}
\label{alm}
\partial_z\partial_{0}\tilde{A}_{0} + \partial_z\left(\frac{G^{ii}}{G^{00}}\right)\partial_{i}\tilde{A}_i + \frac{G^{ii}}{G^{00}}\partial_z\partial_{i}\tilde{A}_i= -\partial_z\left(\frac{G^{ii}}{G^{00}}\right)\partial_{i}\partial_{i} \chi_2.
\end{equation}
The constraint (\ref{constraint}) for this case reduces to
\begin{equation}
G^{00}\partial_z \partial_{0}\tilde{A}_{0} + G^{ii}\partial_z \partial_{i}\tilde{A}_{i}= 0,
\end{equation}
which simplifies equation (\ref{alm}) and yields
\begin{equation}
\partial_{i}\tilde{A}_i = -\partial_{i}\partial_{i} \chi_2,
\end{equation}
which is impossible to satisfy since $\partial_{i}\tilde{A}_i$ is $z$-dependent in general, unlike the right hand side.

We are thus forced to conclude that the radiation gauge choice is impossible to implement for the AdS black hole. For a non-exhaustive list of instances where such has been done, we refer to \cite{Fujita:2009wc,Fujita:2009ca,Grigoryan:2010pj,Cui:2011ag,He:2013qq,Hohler:2013vca,Yang:2015aia}.

\section{Example wherein the inadmissable radiation gauge affects the physical prediction}
\label{odesolving}

The gauge choice issue that we highlighted in the previous section, is relevant for the momentum dependence of the spectral functions. From the latter quantity, we can infer information on the melting behavior of the quarkonium in the plasma. Meson melting in a holographic context was also considered in e.g.~\cite{Fujita:2009wc,Fujita:2009ca,Peeters:2006iu,Ishii:2014paa,Ali-Akbari:2014gia}, and holographic quarkonia in \cite{Fujita:2009wc,Fujita:2009ca,Hohler:2013vca,Grigoryan:2010pj,Hong:2003jm,Kim:2007rt,Hou:2007uk}.  In \cite{Fujita:2009ca} it was argued that, even though a spatial momentum breaks isotropy, the spectral functions still are, quite miraculously, isotropic.\footnote{In \cite{Fujita:2009ca}, this was then compared to the finding of \cite{Hatta:2008tx} that a strongly coupled plasma described by $\mathcal{N}=4$ SYM cannot support jets.} We will demonstrate that this actually arises due to the faulty choice of gauge: if one does not make the additional gauge choice $\partial^{\mu}A_{\mu}=0$, then isotropy is broken as it would be expected when momentum is inserted. \\

\noindent Our main interest here is on the charmonium bound state where we focus on the vector modes ($A_L = A_R$). To start, one makes the Ansatz $A_{\mu}\sim e^{i\mathbf{k}\cdot \mathbf{x} - i \omega t}$ for the dependence on the boundary coordinates. We will choose $\mathbf{k}$ along the 3-axis. \\
\noindent If one includes spatial momentum into the spectral functions, the transverse polarization differential equation for $A_1$ becomes
\begin{equation}
\label{V1par}
\partial^2_{z}A_{1} + \partial_z\left(\ln\left(\sqrt{-\mathcal{G}}e^{-cz^2}G^{zz}G^{11}\right)\right)\partial_z A_1 - \frac{1}{G^{zz}}\left(G^{tt}\omega^2 + k^2G^{33}\right)A_1 = 0,
\end{equation}
where the metric of the black hole (\ref{adsbh}) is to be used. This ODE can be readily solved numerically using the same methods as in \cite{Fujita:2009wc,Fujita:2009ca,Dudal:2014jfa}. We will present the results further on. \\

\noindent The $A_3$ polarization on the other hand becomes much more intricate: it couples directly to $A_0$ and one hence needs to solve a coupled system of differential equations. These are given by
\begin{equation}
\label{ode1}
\partial^2_zA_0 + \partial_z\ln\left|e^{-\phi}\sqrt{-\mathcal{G}}G^{zz}G^{00}\right|\partial_z A_0 - \frac{G^{33}}{G^{zz}}\left(k^2A_0 + k\omega A_3\right) = 0,
\end{equation}
and
\begin{equation}
\label{ode2}
\partial^2_zA_3 + \partial_z\ln\left|e^{-\phi}\sqrt{-\mathcal{G}}G^{zz}G^{33}\right|\partial_z A_3 - \frac{G^{00}}{G^{zz}}\left(\omega^2A_3 + k\omega A_0\right) = 0.
\end{equation}
One readily finds that the combined equation to be solved equals
\begin{align}
\partial_z\left(\frac{G^{zz}}{G^{33}}\partial^2_{z}A_0\right) &+ \partial_z\left(\frac{G^{zz}}{G^{33}}\partial_z \ln\left|e^{-cz^2}\sqrt{-\mathcal{G}}G^{zz}G^{00}\right|\partial_z A_0\right)  -k^2\partial_z A_0 - \omega^2 \frac{G^{00}}{G^{33}}\partial_z A_0 = 0.
\end{align}
This is a second order differential equation for $V= \partial_z A_0$. A Frobenius analysis yields the following asymptotic behavior of $V$. For $z\approx 0$, one finds for the two independent solutions $\Phi_1$ and $\Phi_2$:
\begin{align}
\Phi_1(\epsilon) &= \epsilon, \quad \Phi_1'(\epsilon) = 1, \\
\Phi_2(\epsilon) &= \epsilon \ln(\epsilon), \quad \Phi_2'(\epsilon) = 1 + \ln(\epsilon).
\end{align}
For $z\approx z_h$, one finds
\begin{equation}
\phi_{\pm} \sim \left(1-\frac{\xi}{\xi_h}\right)^{\pm i \omega \xi_h/4}.
\end{equation}
With these boundary values, one can solve the differential equation for $V$ from the boundary $z=0$ to the horizon $z=z_h$. \\
Following the real-time dictionary \cite{Son:2002sd,Policastro:2002se}, see also \cite{Teaney:2006nc,Skenderis:2008dh,Skenderis:2008dg,Lindgren:2015lia}, a linear combination of $\Phi_1$ and $\Phi_2$ has to be taken to satisfy ingoing boundary conditions at the horizon. We parameterize this $V$ as
\begin{equation}
V(z) = C(\Phi_1 + B\Phi_2),
\end{equation}
in terms of two complex numbers $C$ and $B$.
With this solution for $V$, the spectral function can be distilled as follows. Evaluating (\ref{ode1}) near the boundary, one finds
\begin{equation}
\partial_z V(\epsilon) - \frac{1}{z}V(\epsilon) + k^2A_0(\epsilon) +k\omega A_3(\epsilon) = 0,
\end{equation}
or with the above form of $V$:
\begin{equation}
C(1 + B(1+\ln(\epsilon))) - \frac{1}{\epsilon}C(\epsilon + B\epsilon\ln(\epsilon)) + k^2A_0 (\epsilon)+k\omega A_3(\epsilon) = 0,
\end{equation}
yielding
\begin{equation}
C = \frac{k^2A_{0}(\epsilon) + k\omega A_3(\epsilon)}{B}.
\end{equation}
The coupled differential equation leads to three correlators: $\left\langle J_0J_0\right\rangle$, $\left\langle J_0J_3\right\rangle$ and $\left\langle J_3J_3\right\rangle$ where the currents are the charm vector currents. For instance, in evaluating the spectral function for $A_0$, one needs to evaluate the combination \cite{Son:2002sd,Policastro:2002se}
\begin{equation}
  D_R(\omega,k)\sim \lim_{z\to0}\frac{A_0 \partial_z A_0}{z}
\end{equation}
to obtain the retarded Green function, from which follows the spectral function as
\begin{equation}
  \rho(\omega,k)=-\frac{1}{\pi}\Im D_R(\omega,k).
\end{equation}
We therefore look at
\begin{equation}
-\lim_{z\to0}\frac{A_0 \partial_z A_0}{z}
\end{equation}
and functionally differentiate it w.r.t.~the boundary value of $A_0$ twice. This equals
\begin{equation}
-\frac{A_0(\epsilon) \frac{k^2A_{0}(\epsilon)}{B}(\epsilon + B\epsilon \ln(\epsilon)) }{\epsilon} = - A_0(\epsilon)A_{0}(\epsilon) \frac{k^2}{B}(1 + B\ln(\epsilon)) = -A_0(\epsilon)A_{0}(\epsilon) \left(\frac{k^2}{B} +  k^2\ln(\epsilon)\right).
\end{equation}
The second term can be renormalized by introduction of a local counterterm (cf.~holographic renormalization). \\

One can also look at the $A_3 A_0$ correlator, though we will have no interest in this mixed correlator in this work. \\

Finally, we can also look at the $A_3 A_3$ correlator. The correlator can in this case be distilled as (using $\omega G^{00} \partial_z A_0 = k G^{33}\partial_z A_3$)
\begin{equation}
\frac{A_3(\epsilon) \partial_z A_3}{\epsilon} = -A_3(\epsilon) \frac{\omega \partial_z A_{0}( \epsilon)}{k\epsilon} = -\frac{A_3(\epsilon) A_3(\epsilon) \omega k\omega }{k B\epsilon}(\epsilon + B\epsilon \ln(\epsilon))  = -A_3(\epsilon)A_{3}(\epsilon) \left(\frac{\omega^2}{B} +  \omega^2\ln(\epsilon)\right),
\end{equation}
so for the spectral function, it turns out we should look at $-\Im\frac{\omega^2}{B}$. It is this correlator that we will be interested in in the remainder of this work.

\subsection{Results}
Below are the resulting Figures of the spectral functions as $k$ is varied for the two different polarizations. We denote everything in dimensionless quantities (by rescaling $z$) where in particular
\begin{equation}
\xi = \sqrt{c}z, \quad \tilde{\omega} = \frac{\omega}{\sqrt{c}}, \quad \tilde{\mathbf{k}} = \frac{\mathbf{k}}{\sqrt{c}}.
\end{equation}
\begin{figure}[h]
\centering
\includegraphics[width=0.7\linewidth]{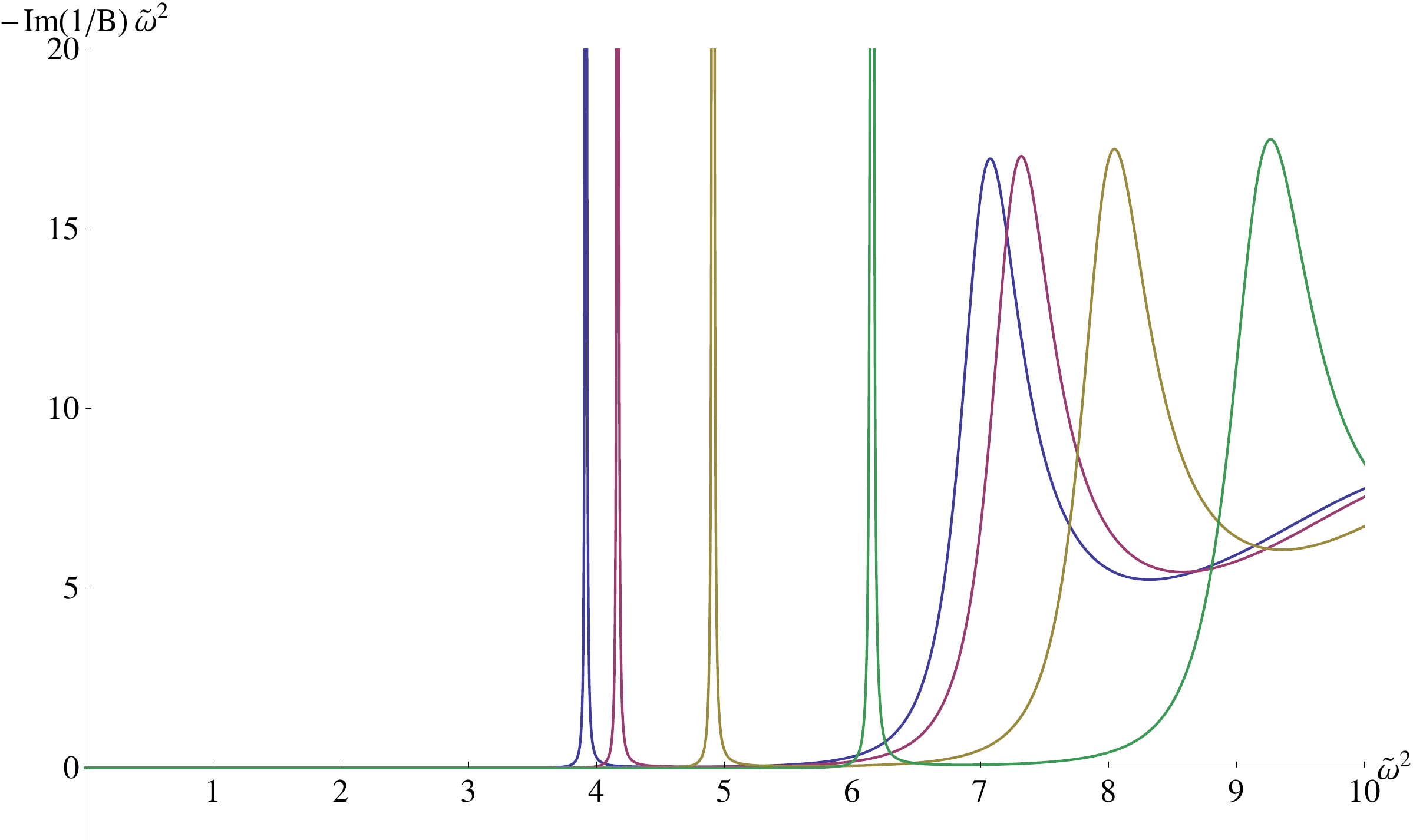}
\caption{Spectral function of $A_3$ for different values of $\tilde{k}$ at $t=0.07$. Blue: $\tilde{k}=0.0$, purple: $\tilde{k}=0.5$, yellow: $\tilde{k}=1.0$, green: $\tilde{k}=1.5$.}
\label{momdep1}
\end{figure}
\begin{figure}[h]
\centering
\includegraphics[width=0.7\linewidth]{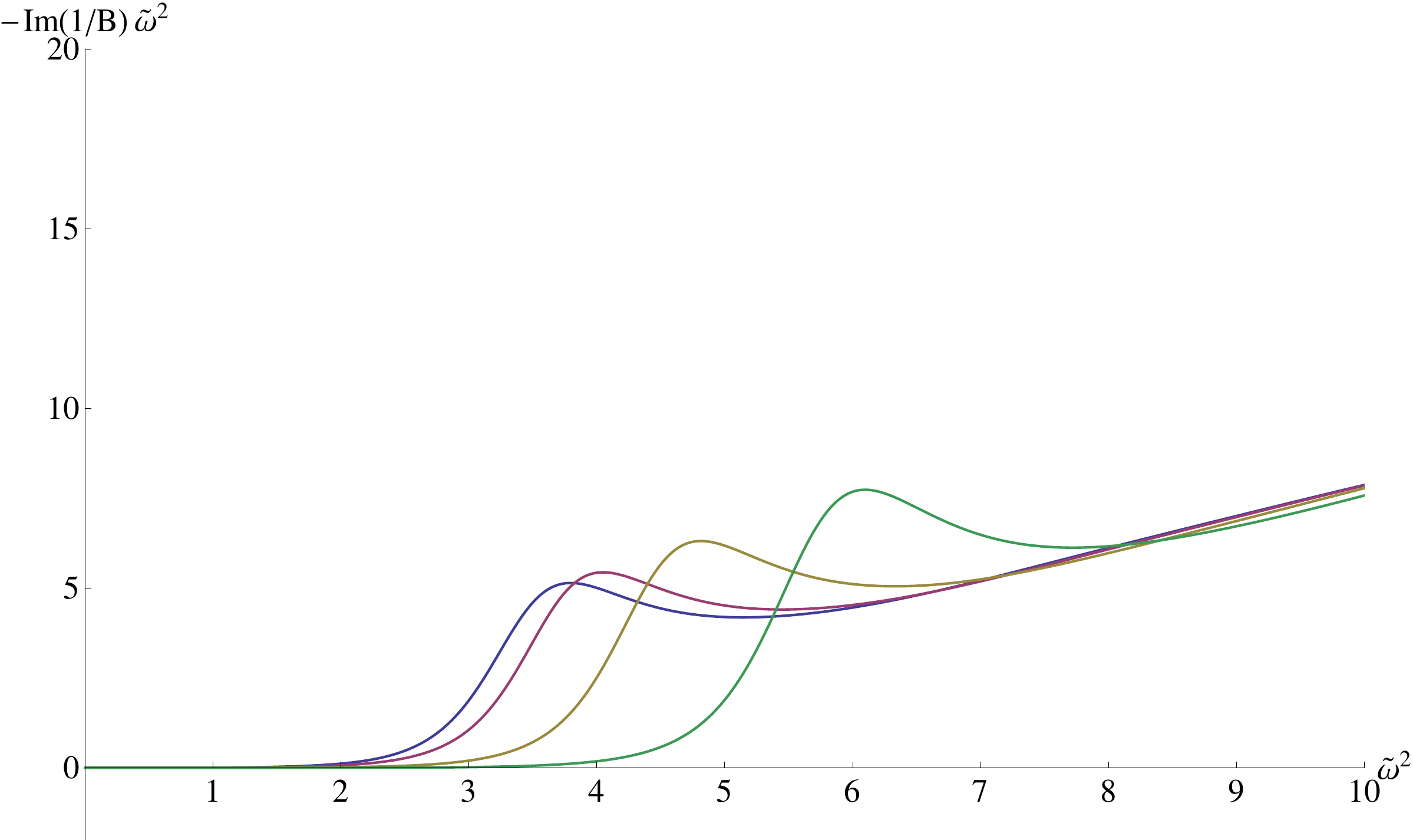}
\caption{Spectral function of $A_3$ for different values of $\tilde{k}$ at $t=0.13$. Blue: $\tilde{k}=0.0$, purple: $\tilde{k}=0.5$, yellow: $\tilde{k}=1.0$, green: $\tilde{k}=1.5$.}
\label{momdep1t}
\end{figure}
\begin{figure}[h]
\centering
\includegraphics[width=0.7\linewidth]{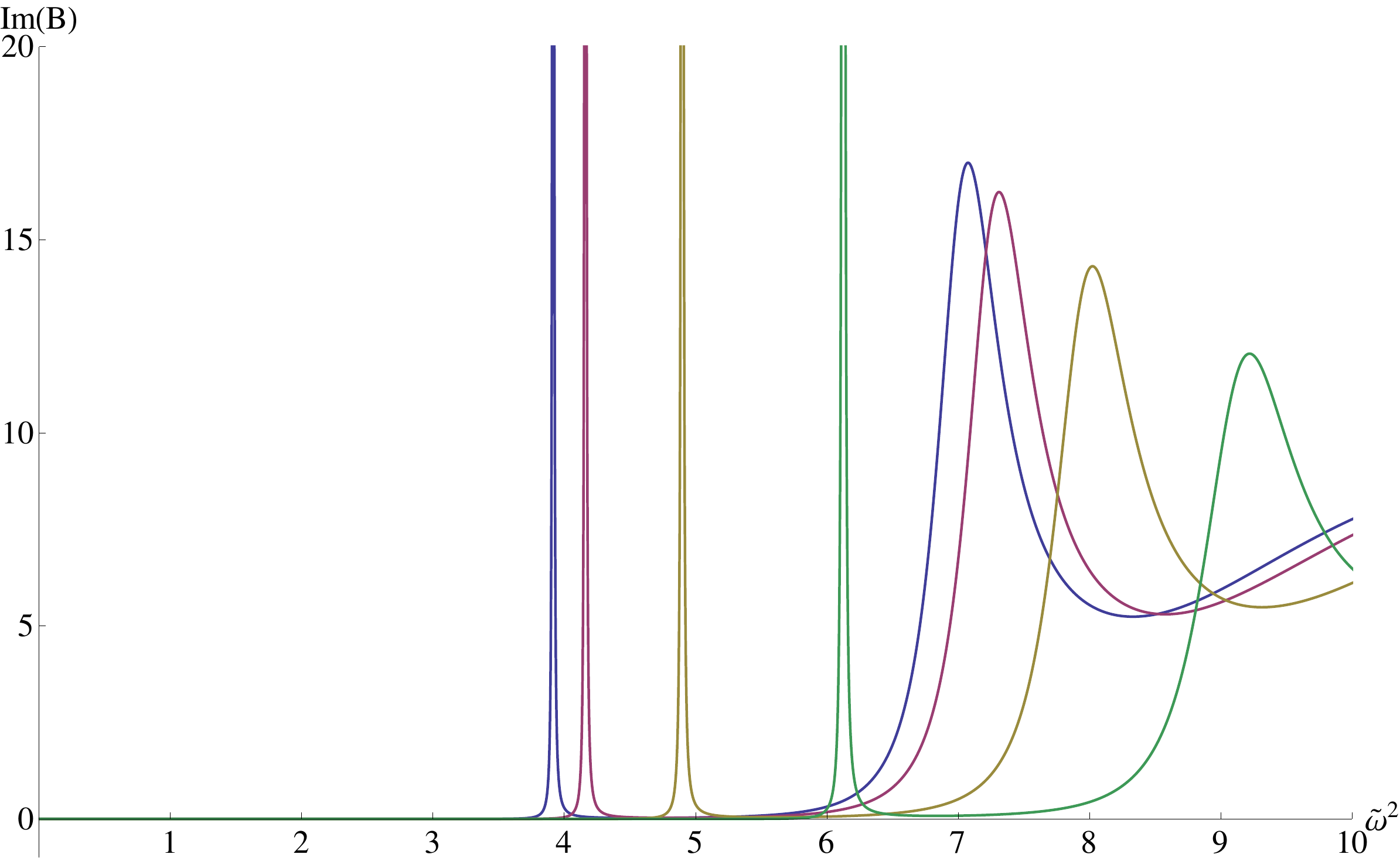}
\caption{Spectral function of $A_1$ for different values of $\tilde{k}$ at $t=0.07$. Blue: $\tilde{k}=0.0$, purple: $\tilde{k}=0.5$, yellow: $\tilde{k}=1.0$, green: $\tilde{k}=1.5$.}
\label{bv1parmom}
\end{figure}
Figures \ref{momdep1} and \ref{momdep1t} show the melting of the spectral peaks as the temperature is increased. This effect remains the same regardless of the value of $k$. Comparing Figures \ref{momdep1} and \ref{bv1parmom}, it is clear that the spectral functions are not isotropic: if the polarization is tangential to the momentum, the peaks hardly decline; whereas if the polarization is perpendicular to the momentum, the peaks decrease quickly as the momentum is increased. Note that for $k=0$, the spectral functions indeed are the same: this is of course expected as in this case $\mathbf{k}=\mathbf{0}$ and hence full isotropy should be restored indeed. \\

\begin{figure}[h]
\centering
\includegraphics[width=0.7\linewidth]{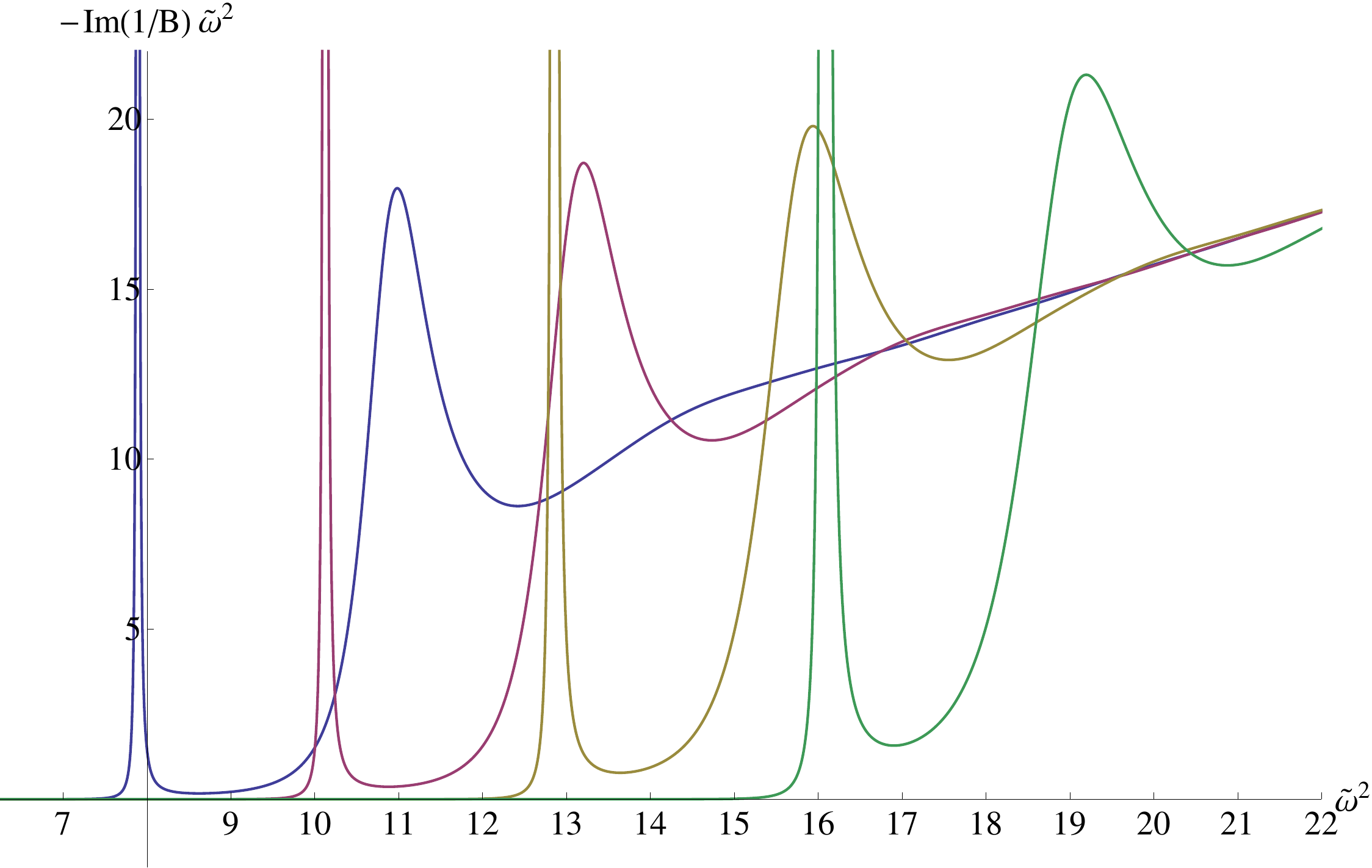}
\caption{Spectral function of $A_3$ for different values of $\tilde{k}$ at $t=0.07$. Blue: $\tilde{k}=2.0$, purple: $\tilde{k}=2.5$, yellow: $\tilde{k}=3.0$, green: $\tilde{k}=3.5$.}
\label{momdepf}
\end{figure}

Figure \ref{momdep1} has the peculiar property that the spectral peaks do not damp as $k$ increases. If one increases $k$ even further, one finds the result of Figure \ref{momdepf}.

\section{Influence of magnetic field on the spectral functions}
In this section, we will demonstrate how the spectral functions are modified if a background magnetic field $\mathcal{B}$ is turned on. In order to analyze the influence of the magnetic field, we will follow our earlier model \cite{Dudal:2014jfa} where we proposed a DBI-modification of the soft-wall model to allow the magnetic field to couple to the charged constituents of the mesons. If one compares the differential equations to those described in the previous section, only a few things change. The background $\mathbf{\mathcal{B}}$-field, which we orient along the 3-axis, corresponds to
\begin{equation}
\bar{F}_{12} = - \bar{F}_{21} = \partial_1 A_2 = -iq\mathcal{B}\frac{2}{3},
\end{equation}
since the charm quark charge is $+2/3 q$. We obtain for $\mathcal{G}_{\mu\nu} \equiv g_{\mu\nu} + 2\pi\alpha' i F_{\mu\nu}$:
\begin{eqnarray}
\mathcal{G}_{\mu\nu}=\left[\begin{array}{ccccc}
g_{00} & 0 & 0 & 0 & 0\\
0 & g_{11} & 2\pi\alpha' i \bar{F}_{12} & 0 & 0  \\
0 & -2\pi\alpha' i \bar{F}_{12} & g_{22} & 0 & 0\\
0 & 0 & 0 & g_{33} & 0 \\
0 & 0 & 0 & 0 & g_{zz} \end{array}\right],
\end{eqnarray}
with its determinant
\begin{equation}
\mathcal{G} = g_{00}g_{33}g_{zz}\left(g_{11}g_{22} - (2\pi\alpha')^2\bar{F}_{12}^{2}\right).
\end{equation}

We will denote by $G$ only the symmetric part of the metric tensor $\mathcal{G}$:
\begin{eqnarray}
G^{\mu\nu}=\left[\begin{array}{ccccc}
\frac{1}{g_{00}} & 0 & 0 & 0 & 0\\
0 & \frac{g_{22}}{X} & 0 & 0 & 0  \\
0 & 0 & \frac{g_{11}}{X} & 0 & 0\\
0 & 0 & 0 & \frac{1}{g_{33}} & 0 \\
0 & 0 & 0 & 0 & \frac{1}{g_{zz}} \end{array}\right],
\end{eqnarray}
where $X = g_{11}g_{22} - (2\pi\alpha')^2\bar{F}_{12}^{2}$. \\

Then substituting these values of $\sqrt{\mathcal{G}}$ and $G^{\mu\nu}$ in the earlier differential equations (\ref{V1par}), (\ref{ode1}) and (\ref{ode2}), one obtains the correct equations for the DBI-modification of the model.\footnote{The string length parameter $\alpha'= \ell_s^2$ was fixed in \cite{Dudal:2014jfa} in terms of the AdS length $L$. The latter only figures as an overall prefactor of the action and we ignore it further on.} Moreover, the asymptotic behavior of the solutions both at the horizon and at the boundary is unaffected by including the magnetic field. \\

\begin{figure}[h]
\centering
\includegraphics[width=0.2\linewidth]{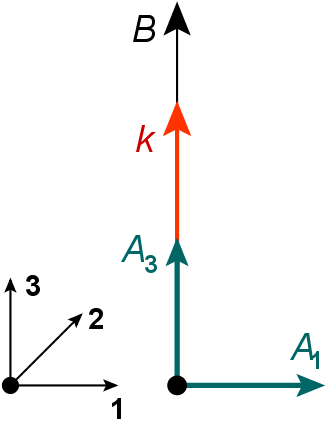}
\caption{Set-up: the magnetic field is directed along the 3-axis, the momentum is chosen to be tangential to the magnetic field and the two polarizations that we will consider are along the 1- and the 3-axis.}
\label{configur}
\end{figure}

For the sake of brevity, we will only look at the special case where the spatial momentum $\mathbf{k}$ is tangential to the magnetic field $\mathbf{\mathcal{B}}$. The other case can be dealt with analogously but will not be discussed here. The set-up of the different vector quantities is demonstrated in Figure \ref{configur}.

\begin{figure}[h]
\centering
\includegraphics[width=0.7\linewidth]{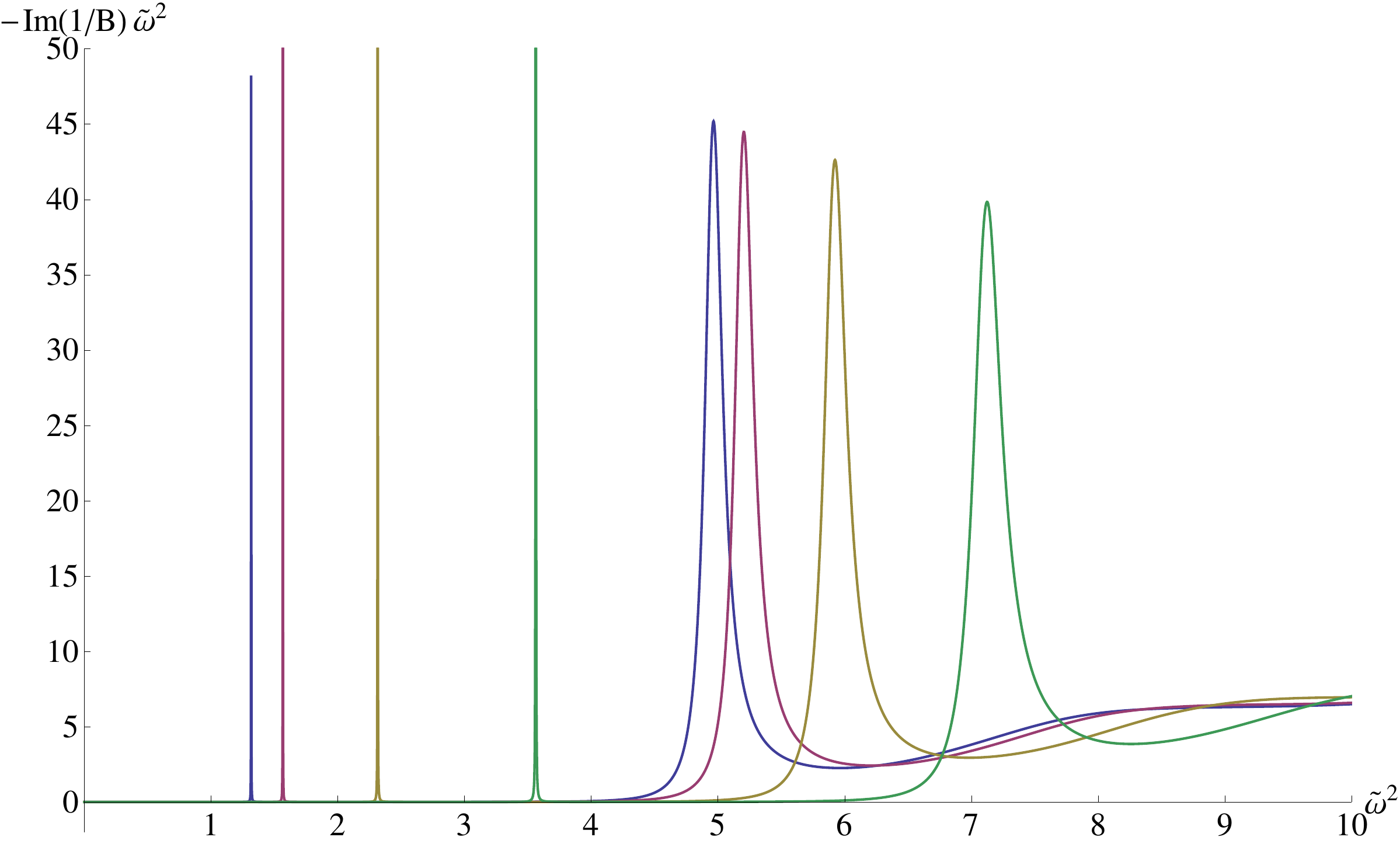}
\caption{Spectral function of $A_3$ with $q\mathcal{B}=1.0$ GeV$^2$ for different values of $\tilde{k}$ at $t=0.07$. Blue: $\tilde{k}=0.0$, purple: $\tilde{k}=0.5$, yellow: $\tilde{k}=1.0$, green: $\tilde{k}=1.5$.}
\label{momdep2}
\end{figure}

\begin{figure}[h]
\centering
\includegraphics[width=0.7\linewidth]{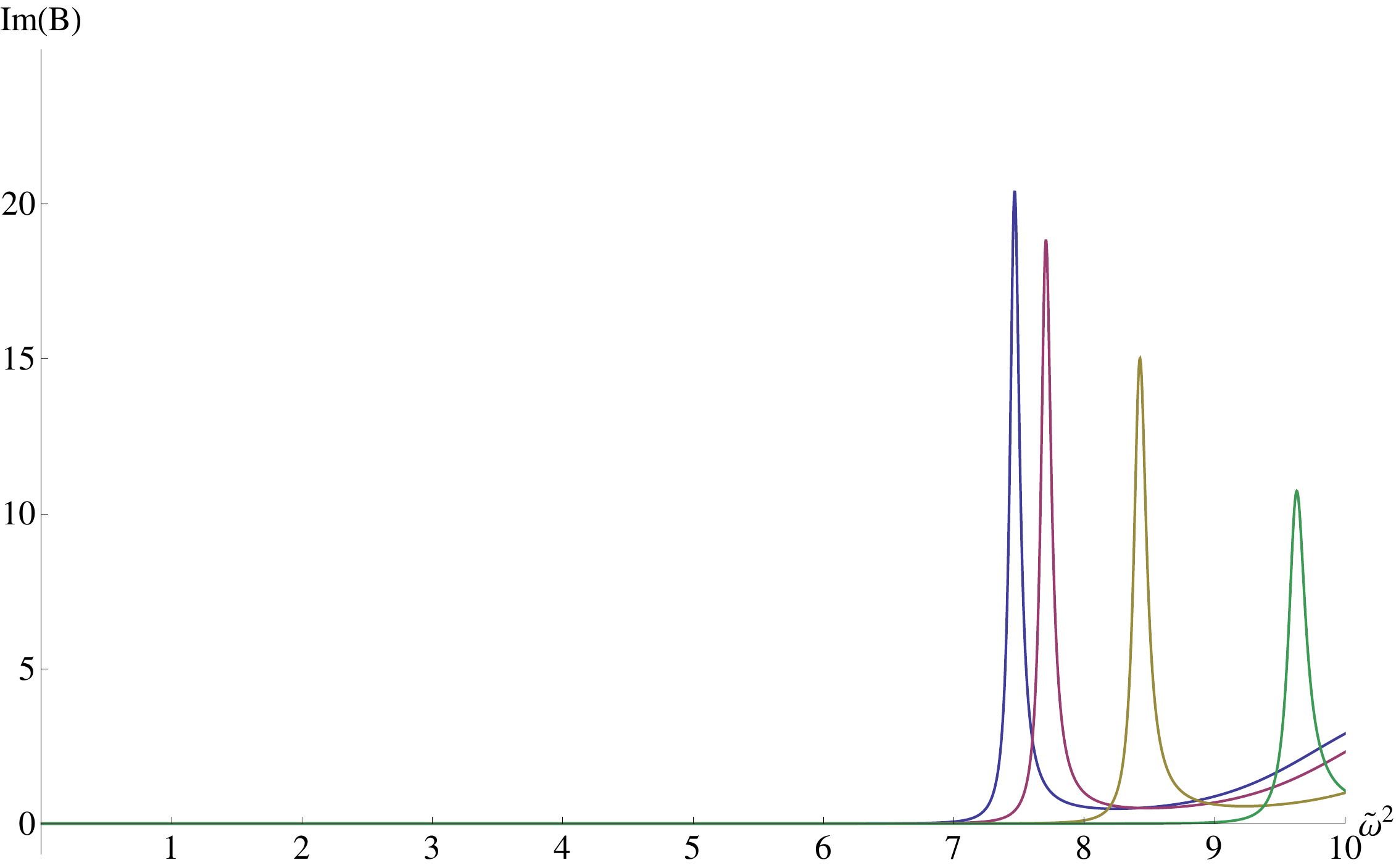}
\caption{Spectral function of $A_1$ with $q\mathcal{B}=1.0$ GeV$^2$ for different values of $\tilde{k}$ at $t=0.07$. Blue: $\tilde{k}=0.0$, purple: $\tilde{k}=0.5$, yellow: $\tilde{k}=1.0$, green: $\tilde{k}=1.5$.}
\label{bv1parmomB}
\end{figure}

Turning on the magnetic field, one finds the results of Figures \ref{momdep2} and \ref{bv1parmomB}.
First of all, we note that the spectral functions are not the same, even for $k=0$: this corresponds to the breaking of the isotropy already by the magnetic field alone.

Of course, we have only discussed the case where $\mathbf{k}$ and $\mathbf{\mathcal{B}}$ are aligned. The other case where for instance $\mathbf{k}$ is directed along the 1-axis can also be studied analogously, though we will not pursue this here. \\

Both the results of this and the previous section demonstrate that turning on momentum causes the peaks to shift to higher $\omega$ as expected, where the peak location is at fixed $\omega^2-k^2$. The reason for this location is that these peaks are identified with the delta-peaks in the thermal AdS case (where $f=1$). In that case, the time direction and the spatial directions along the boundary are fully equivalent (no warping in the $z$-direction) and full Lorentz invariance should be manifestly present.

As a conclusion, all spectral functions widen as $k$ is increased; this implies the excitations melt at a lower temperature in correspondence with \cite{Fujita:2009wc,Fujita:2009ca}: the meson melts under the hot wind of the quark-gluon plasma. We do want to remark that the height of the spectral peak on the other hand does not systematically decrease, but this depends on the polarization and the magnitude of the magnetic field. \\

\section{Conclusion}
We have scrutinized the quite common application of the radiation gauge in holography and demonstrated that for the AdS black hole, one cannot impose this gauge. We then demonstrated that this issue is of physical relevance.  This was achieved by showing that the previously observed emergent isotropy for the momentum-dependent (quarkonium) spectral function is actually the result of a faulty choice of gauge. As the momentum $k$ increases, the spectral peaks widen and melt at a lower temperature than before. We furthermore provided some results on the momentum dependence of the spectral functions when including a background magnetic field, following our earlier proposed model \cite{Dudal:2014jfa} of a DBI-extension of the soft wall model.

\section*{Acknowledgments}
We thank D.~R.~Granado for collaboration at an early stage of this work. T.~Mertens gratefully acknowledges financial support from the UGent Special Research Fund, Princeton University, the Fulbright program and a Fellowship of the Belgian American Educational Foundation.

\end{document}